# 'Seed+Expand': A validated methodology for creating high quality publication oeuvres of individual researchers

Linda Reijnhoudt[1], Rodrigo Costas[2], Ed Noyons[2], Katy Börner[1,3], Andrea Scharnhorst[1]

[1] *linda.reijnhoudt@dans.knaw.nl, andrea.scharnhorst@dans.knaw.nl*
DANS, Royal Netherlands Academy of Arts and Sciences (KNAW), the Hague, the Netherlands

[2] *rcostas@cwts.leidenuniv.nl, noyons@cwts.leidenuniv.nl*
Center for Science and Technology Studies (CWTS)-Leiden University, Leiden, the Netherlands

[3] *katy@indiana.edu*
Cyberinfrastructure for Network Science Center, School of Library and Information Science, Indiana University, Bloomington, Indiana, United States of America

**Abstract**
The study of science at the individual micro-level frequently requires the disambiguation of author names. The creation of author's publication oeuvres involves matching the list of unique author names to names used in publication databases. Despite recent progress in the development of unique author identifiers, e.g., ORCID, VIVO, or DAI, author disambiguation remains a key problem when it comes to large-scale bibliometric analysis using data from multiple databases. This study introduces and validates a new methodology called seed+expand for semi-automatic bibliographic data collection for a given set of individual authors. Specifically, we identify the oeuvre of a set of Dutch full professors during the period 1980-2011. In particular, we combine author records from the National Research Information System (NARCIS) with publication records from the Web of Science. Starting with an initial list of 8,378 names, we identify 'seed publications' for each author using five different approaches. Subsequently, we 'expand' the set of publications in three different approaches. The different approaches are compared and resulting oeuvres are evaluated on precision and recall using a 'gold standard' dataset of authors for which verified publications in the period 2001-2010 are available.

**Conference Topic**
Old and New Data Sources for Scientometric Studies: Coverage, Accuracy and Reliability (Topic 2) and Management and Measurement of Bibliometric Data within Scientific Organizations (Topic 9)

**Introduction**
Creating correct linkages between a unique scholar authoring a work and her or his (possibly many) author name(s) is complex and unresolved. Authors might use anonymous and alias author names, names might be misspelled or change over time, e.g., due to marriage, and multiple scholars might have the very same name. Yet, science is driven by scholars, and the identification and attribution of works to individual scholars is important for understanding the emergence of new ideas, to measure the creative human capital of institutions and nations, to model the relationships and networks of researchers, and to forecast new scientific fields (Scharnhorst et al. 2012). With the 'return of the author' in bibliometrics (Scharnhorst & Garfield 2011); bibliometric indicators on the individual level (Hirsch, 2005; Costas et al, 2010; Lariviere, 2010; Vieira & Gomes, 2011); and institutional evaluation based on the individual publication output of authors over longer time periods (van Leeuwen, 2007; Zuccala et al, 2010), the ambiguity problems in allocating publications to authors have become more pressing (Costas et al, 2005, 2010). Different approaches to data collection at the individual level have been proposed in the literature (see the review by Smalheiser & Torvik, 2009), although in many cases these approaches focus on the disambiguation of author in one single database (e.g. PubMED). Recently, systems of unique author identifiers offer a practical solution, e.g. ORCID. However, they are not yet fully standardized and often rely on authors to register their own bibliographic profiles. Thus, the problem of automatically linking author names across publication, patent, or funding databases still persists.

In this paper, we present a general methodology that combines information from different data sources[1] to retrieve scientific publications covered in the Web of Science (WoS) for a given list of authors. Specifically, we trace the publications for 8,378 professors affiliated with at least one of the Dutch universities, as included in the Nederlandse Onderzoek Databank (NOD, Dutch Research Database) and displayed in the web portal NARCIS (National Academic Research and Collaborations Information System). The approach differs from prior work by the usage of an initial set of 'seed publications' for each author; and the expansion of this seed to cover the whole oeuvre of each author as represented in a large bibliographic database using an automated process. This automatic process is applied in parallel to each of the authors in the initial set. At the end, an ensemble of authors and their publications is build from the individual oeuvres. We compare five different approaches to create 'seed publications' and three approaches to 'expand' the seed. Last but not least, we assess and validate the proposed methodology against a 'gold standard' dataset of Dutch authors and their publications that was compiled by Centre for Science and Technology Studies (CWTS) and verified by the authors themselves.

The proposed methodology is able to account for different kinds of author ambiguity such as different ways of spelling a name, different ways to store a name (initials, first and last name, etc.) in different database systems, and misspellings. Homonyms, i.e., different authors with the same name, can be partially resolved using additional information about authors such as address and institution information yet, two authors might work at the same institution and on similar topics and be merged. Note that each new information source likely offers new challenges. Plus, there is much human error that is hard or impossible to detect: Mail addresses can be wrongly noted or allocated, even publications verified by the authors themselves can be wrong. Because we aim for a scalable methodology of automatic oeuvre detection we counter these ambiguities by different means:

- manual cleaning of initial sets for automatic retrieval
- manual inspection of multiple links in automatically produced mappings
- applying similarity measures (as Levenshtein distance) for string comparison
- elimination of most common names and
- elimination of publications when more than one professor from the sample with similar names are matched to the same author in a given paper (e.g. Ad de Jong and Albert de Jong assigned to the same paper with the author "A. de Jong").

The rest of the paper is organized as follows. First, a description of the different datasets is given, followed by an overview of the 'seed creation' approach applied. Second, the 'expansion of the seed' and the results and issues of performance are presented. Finally, we discuss the key results of the proposed methodology, draw conclusions, and discuss planned work.

**Data**

The datasets used in this study are under active development at DANS (Data Archiving and Networked Services), an institute of the Royal Netherlands Academy of Arts and Sciences (KNAW) and the Centre for Science and Technology Studies (CWTS). DANS promotes sustained access to digital research data and also provides access, via NARCIS.nl, to thousands of scientific datasets, e-publications and other research information in the Netherlands. In addition, the institute provides training and advice, and performs research into sustained access to digital information. CWTS is a centre of excellence in bibliometric analysis. It has conducted numerous bibliometric studies both for research and for evaluation, and compiled extensive data about Dutch researchers.

*NARCIS/NOD database: The Dutch full professor seed*

KNAW serves the NARCIS Dutch research information system (Baars et al, 2008) a web portal for a set of databases. One of them is the so-called NOD (Dutch research database) which contains information about forty thousand plus personnel employed at Dutch research institutions (universities and other academic institutions). The person database contains metadata such as names, e-mail addresses, but also—for some scholars among them—the Dutch Digital Author Identifier (DAI). Introduced in 2008 in the Netherlands, the DAI assigns a unique identifier to every employee of a Dutch university, university of applied sciences (HBO-*Hoger beroepsonderwijs*) or research institute. Since 2006, NARCIS also harvests publications from Dutch scientific repositories. These are matched to the scholars on their DAI. A complete dump of the NARCIS database was made on April 3, 2012 and is used in this paper (Reijnhoudt et al, 2012). Specifically, we will use the set of 8,378 *hoogleraren*, or *full professors,* and their 105,128 papers to exemplify the proposed methodology. 75% of the full professors have a known DAI.

*CWTS Web of Science database: High quality publication data*

The in-house CWTS version of the Thomson Reuters Web of Science (WoS) consists of nearly 35 million scientific publications and hundreds of millions of citations, from 1980 up to 2012, covering all fields of science. It comprises the Science Citation Index Expanded (SCIE) as well as different enhancements made during the scientific and commercial activities of CWTS over more than 20 years. Enhancements include among others: The standardization of different fields, namely addresses, journal names, references and citation matching, and a new disciplinary classification at the paper level (Waltman & van Eck, 2012). The methodology proposed here uses the standardized address information and the new classification.

*CWTS SCOPUS database: Scopus Author Identifier*

Scopus is one of the largest abstract and citation databases of peer-reviewed literature. The database contains 47 million records, 70% with abstracts from more than 19,500 titles from 5,000 publishers worldwide covering the years 1996 – 2012 (http://www.info.sciverse.com/scopus/about). Of particular interest for this study is the newly introduced 'Scopus Author Identifier' that is based on an assignment of documents to authors determined by their similarity in affiliation, publication history, subject, and co-authors (Scopus, 2009). It has been discussed that articles assigned to a particular Scopus Author Identifier tend to be articles of the author represented by that identifier, but the set of articles might be incomplete, or articles by the same author might be assigned to multiple identifiers (see Moed et al, 2012).

*CWTS Gold Standard dataset: High quality publication oeuvres*

Frequently CWTS' studies at the individual author level require a manual verification process in which the individual researchers check and verify their own lists of publications - for more details on this verification process see (van Leeuwen, 2007). This verification process has been applied to different sets of researchers in the Netherlands on publications from 2001 to 2010. From this dataset of verified author-publication oeuvres we retrieve a 'gold standard' dataset by manual matching on initials, last names and organisations. This dataset consists of 1400 full professors captured in NARCIS to evaluate different author disambiguation methods. The use of a gold standard set is a common approach in bibliometric and information retrieval research (Costas & Bordons, 2008; Sladek et al, 2006).

## Methodology

The main objective of this study is to develop and validate a general methodology, called seed+expand, for automatic oeuvre detection at the individual author level. Given a set of author names, we are interested to detect their publications, as many (high recall) and as correctly (high precision) as possible combining data from different databases. To exemplify and evaluate the methodology, we use the set of all 8,378 full professors included in the NARCIS database. Figure 1 shows an overview of the workflow comprising:

1. **Seed Creation**: Starting with an initial list of 8,378 full professors, we collect information on their name, affiliation, and e-mail addresses from the NOD. Next, we identify 'seed publications' in the WoS for each author using five different approaches.
2. **Seed Expansion**: Retrieval of additional papers for seed authors based on characteristics of the papers. Three different approaches are compared.
3. **Evaluation**: Results of the different seed expansion approaches are validated using standard measures for precision and recall and the CWTS 'gold standard' dataset of authors for which verified publications in the period 2001-2010 are available.

All three parts are detailed subsequently.

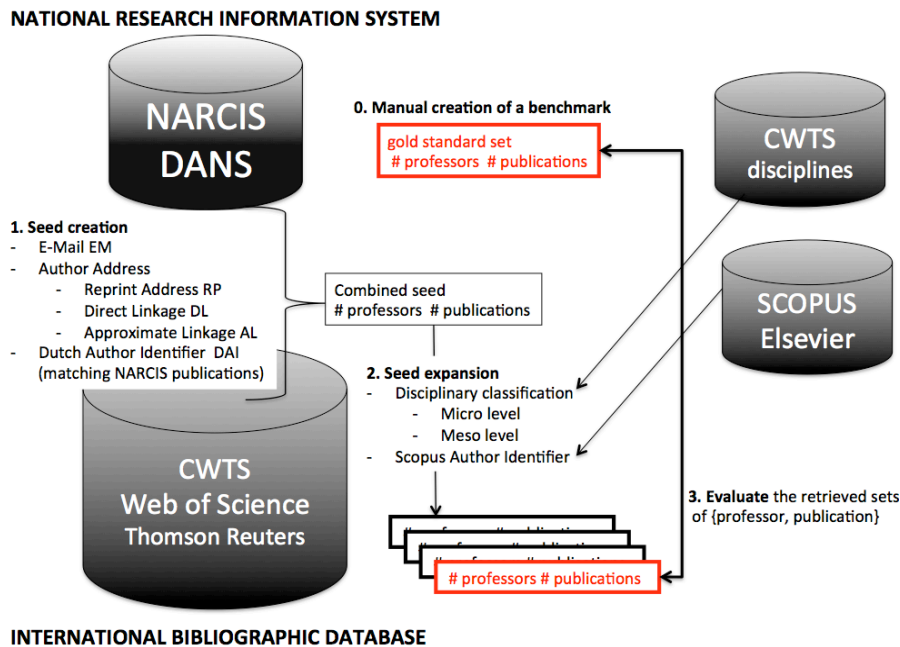

Figure 1: General workflow and relevant data sources

## Seed Creation

The first step of the methodology consists of the creation of a reliable 'seed' of publications for the 8,378 target professors. An element of this set consists of a triplet of elements (publication identifier, person identifier and author position in the paper). The identifiers come from different databases; and the author position indicates if the scholar is first, second, or $n$th author for that publication. The accuracy of the seed is very important as the precision and recall of the final oeuvre detection will be significantly higher if the precision of the seed is high. It is important to bear in mind that during the expansion phase of the methodology it will not be possible to add papers for those professors that are not already in the seed. So, for this phase in the methodology it is not the recall on papers but the recall on authors that matters. Five different approaches of creating 'accurate' seeds are explored here:

- E-mail seed (EM): A seed based on the matching of the e-mail of the professor with the publications in Web of Science.
- Three author-address approaches (RP, DL, AL): These seeds are based on different combinations of the name of the professors and the affiliation(s) of that professor matched with the Web of Science.
- DAI seed (DAI): This approach builds upon the publications in NARCIS that have been attached to the professors through the Dutch Author Identifier.

*E-mail Seed (EM)*

In the NOD system, e-mail addresses are attached to scholars directly or via their affiliations. Hereby, e-mail addresses of the target professors are simply matched against e-mail addresses of authors found in the papers in Web of Science. This approach produced a seed for 4,786 different authors (57% of the professors from our list) with at least one paper found in WoS, see also Table 1. As e-mail addresses are uniquely attached to one scholar[2] and are seldom transferred to other scholars[3], this approach is assumed to be most accurate.

*Three Author-Address approaches*

Three approaches combine author names and affiliation data in the NOD system and match them to WoS affiliation data to retrieve relevant publications. This approach was only feasible thanks to standardization of WoS affiliations and addresses by CWTS. However, some parts of this task also required manual handling and checking. As a result of this, 92% of the papers with a Dutch organization in the WoS have a matched counterpart in the NOD organizations. These are the only 652,978 papers that can be considered for the Author-Address approach seeds as described in the following paragraphs.

*Reprint author (RP):* In scientific publications, the reprint address refers to the address of the corresponding author in charge of managing requests that a publication may generate. In the WoS database this reprint address appears directly linked to the author, thus offering a direct and "safe" connection between an author and an organization that can be directly extracted from the publication. Thus, the creation of this seed consists of the matching of the name of the professor and his/her affiliation as is recorded in the NOD with the reprint author name and the reprint affiliation.

*Direct linkage author-addresses (DL):* 69% of publications WoS include data on the linkage between the authors and their organizations as they appear in the original publications. For instance, if the original publication featured three authors and two organizations this linkage of authors and organizations is indicated as follows (Figure 2):

> Author A(1)(2), Author B(2), Author C(1)
> Organization G, Organization H

Figure 2: Example of direct author organization linkage

Indicating that Author A is linked to Organization G and H; Author B is linked to Organization H, and Author C is linked to Organization G. As in the RP-based approach, the names and the affiliations of the professors are matched with the author-affiliation linkages of the publications, detecting those publications that, based on this author-affiliation linkage, could belong to the target professors.

*Approximate linkage author-addresses (AL):* The other 31% of publications did not have a direct linkage recorded in the database. Thus, authors and affiliations of the publications were recorded, but there was no way to tell which author is affiliated with what organization. This approach detects as seed publications of the target professor those publications that share the same name and affiliation as the professor in the same record. The AL approach has the potential problem of wrongly attributing a publication to a target professor if the name of the author and the institute both appear on a paper. For instance, referring back to Figure 2, if a homonym of 'Author B' (i.e., another scholar with the same name) appears in a paper where 'Organization H' also appears, the real 'Author B' might get this paper wrongly attributed to him/her.

*DAI Seed (DAI)*

This seed creation approach starts with publications that are in the NARCIS database attributed by means of the DAI to the target professors. However, these publications are not necessarily WoS publications (there may be books, theses, or journal articles not covered in the WoS). For this reason, it was necessary to perform a matching process between the bibliographic records for the professors with a DAI extracted from NARCIS and the WoS database. The NARCIS papers were matched with the WoS publications on journal, year, title, and first page. This way, we were able to create a new seed, based on the publications covered in the Web of Science that were also in the NARCIS database for the target professors.

*Combining the seeds*

Table 1 shows the resulting numbers of publications and professors created by the five different seed creation approaches. Publications are counted once per seed, even if they appear several times for different professors. The last column shows the number of professors that were found exclusively by this particular seed method. So, if the AL approach would not be used, the number of professors found would drop only by 76, whereas not using the EM approach would result in a drop of 790 professors. At the end, all seed results are combined (added) and cleaned for duplicates leaving us with 6,989 unique professors and corresponding 174,568 publications.

**Table 1: Result sets obtained by different seeds**

| Seed Method | CWTS Publications | NARCIS Full Professors | Full Professors Unique to This Seed |
|---|---|---|---|
| EM | 40,826 | 4,786 | 790 |
| RP | 81,079 | 5,819 | 149 |
| DL | 79,515 | 5,749 | 158 |
| AL | 28,837 | 5,018 | 76 |
| DAI | 30,322 | 2,742 | 162 |
| **Total unique in combined seed** | 174,568 | 6,989 | |

To further improve seed quality, we remove multiple assignments and common names. Multiple assignments refer to the cases where more than one professor is matched to the same paper, with the same author position number. Clearly this is wrong, as only one researcher should be matched to one author. In order to keep the seed as precise as possible and thus sacrificing some recall for precision, all these records have been removed (see Table 2, 'remove multiple assignments'). The top 5% most common author names (first initial-last name pairs) from the former two seed-approaches (RP and DL) and the top 10% from the least precise approach (AL), thus trying to keep the level of 'noise' (i.e., false positives) in the seed to a minimum. (See
Table 2, column 'remove common names').
The resulting seed comprises 6,753 professors (80% of initial set of 8,378) with 157,343 unique papers.

Table 2: Pruning the seeds to increase precision

| Seed Method | Number of Found Professors | Remove Multiple Assignments | Remove Common Names |
|---|---|---|---|
| EM | 4,786 | 4,786 | 4,786 |
| RP | 5,819 | 5,696 | 4,648 |
| DL | 5,749 | 5,629 | 4,675 |
| AL | 5,018 | 4,864 | 3,147 |
| DAI | 2,742 | 2,742 | 2,740 |
| **Total unique in combined seed** | **6,989** | **6,947** | **6,753** |

**Seed Expansion**

In this second phase, we use the 6,753 author profiles and associated papers in the seed to identify additional publications by these authors in the WoS database. Three approaches have been explored and are detailed subsequently.

*Two CWTS Paper-Based Classifications (Meso and Micro)*

These two approaches use a new paper-based classification that has been developed at CWTS (Waltman & van Eck, 2012) based on the citation relationships of individual publications. It has been applied to publications between 2001 and 2011, excluding the Arts and Humanities. The hierarchical classification has three levels, with a medium-level classification that comprises 672 'meso-disciplines', and a lower-level classification that includes more than 20,000 different 'micro-disciplines.' We assume that within small disciplinary clusters (meso and micro) there is a rather low probability that two professors share the same name. Hence, we assign all papers within a meso/micro-discipline that have the same author names to one professor. Incorrect assignments might occur when two persons with the same name work in the same subfield.
Performing assignments at the meso-discipline level results in an increase of 34% to 211,202 unique papers, the micro-disciplines yield a subset thereof with 194,257 unique papers, an increase of 23%.

*Scopus Author Identifier Approach*

A third approach to expand the publications of professors is to use one of the already existing author identifiers. We choose the Scopus Author Identifier because it has been introduced for all authors in the Scopus database. Here, the 157,343 WoS publications from our initial seed were matched to Scopus publications and their 6,753 authors were matched with Scopus authors to derive their Scopus author identifier. As shown in Figure 3, 614 WoS seed authors had no Scopus author identifier; 2,977 authors had exactly one Scopus author identifier; and all others had more than one. All Scopus author identifiers were used to retrieve additional Scopus publications that were traced back to WoS publications via bibliographic matching on journal, title, etc. The resulting set has 266,105 unique papers, an increase by 69%—the largest number of all three approaches.

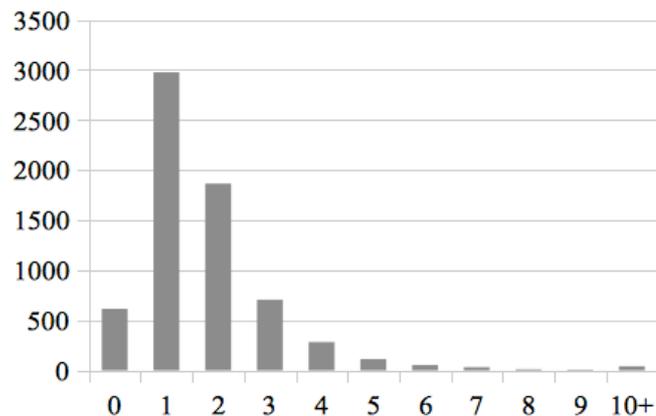

**Figure 3: Number of authors (y-axis) from the seed with a given number of matched Scopus author identifiers (x-axis). The 47 authors with more than 10 were ultimately discarded.**

**Evaluation**

To evaluate the three seed expansion approaches, their result sets are compared to the CWTS gold standard dataset introduced in section 2.4. The expansion of the seed by the different approaches has been performed on the whole WoS (from 1980 to 2011). But to evaluate the approaches we restrain the result of the expansion to publications published between 2001 and 2010, the same time period as the gold standard set. Exactly 1,400 of the 6,753 authors (21%) are in the gold standard dataset - only 63 professors are not accounted for. These 1,400 authors and their 57,775 associated papers will be used to measure precision and recall achieved by the different approaches.

Precision and recall are widely used to measure how well an information retrieval process performs. Precision is defined as the retrieved relevant records (true positives) divided by all retrieved records (both true and false positives). Recall on the other hand is the number of retrieved relevant records divided by the number of records that should have been retrieved (the true positives and the false negatives). Thus, we can score the performance of our different approaches according to these two parameters, see Table 3.

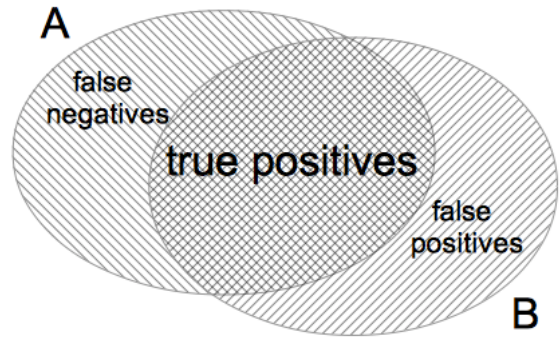

Figure 4: Gold standard set (A) versus the result of the expansion (B)

Column 2-4 in Table 3 present the three approaches individually. In general, the values for recall are equally high for the three approaches. Regarding the precision there is a slight difference. As expected, the micro-discipline set has a higher precision and lower recall than the meso-disciplines approach. The Scopus author identifier approach, with 63,460 professor-paper combinations and a precision of 87.3, ends up exactly in between.

Table 3: Performance of the three expansion approaches - individually and combined

|  | Scopus Identifier | Meso | Micro | ScopusI & Meso | ScopusI & Micro |
| --- | --- | --- | --- | --- | --- |
| True pos. (A∩B) | 55,405 | 55,459 | 55,394 | 55,509 | 55,460 |
| False pos. (¬A∩B) | 8,055 | 10,430 | 7,212 | 13,200 | 10,260 |
| False neg. (A∩¬B) | 2,370 | 2,316 | 2,381 | 2,260 | 2,315 |
| Precision | 87.3 | 84.2 | 88.5 | 80.8 | 84.4 |
| Recall | 95.9 | 96.0 | 95.9 | 96.1 | 96.0 |

The last two columns show the combination of two approaches: *Scopus author id plus meso-disciplines* and *Scopus author id plus micro-disciplines*. As can be expected, the recall increases, whereas the precision declines. The increase in the recall is rather small, and the number of false negatives is high. This indicates that both approaches miss roughly the same papers. Apparently there are some publications in the oeuvres of some researchers that are hard to find using the kind of approaches presented in this paper.

**Conclusions**

Of the 8,378 professors in our target list we identified at least one publication for 6,753 (80%) professors, which gave us a seed into their oeuvre (as far as covered by the WoS). Combining all publications of all professors we started with a set of 6,753 authors and 157,343 unique publications. After the expansion of the individual publication seeds with the Scopus author id approach and the micro-disciplines approach we find the same recall on the gold standard dataset, and a comparable precision, as shown in

Table 3. The Scopus author id approach finds more unique papers (266,105 vs. 194,257). This can be attributed to the restrictions on the disciplines classification on period (2001-2011) and subject, e.g., the Arts and Humanities WoS publications are not included (Waltman & van Eck, 2012).

The gold standard dataset used for evaluation covers 1,400 (or 21%) of the 6,753 professors and we assume the precision and recall results can be extrapolated to the entire data collection. It is important to remark that the results of precision are slightly conservative due to the fact that for some authors some publications were still missing in their verified set of publications. This happens particularly with authors with high numbers of publications, in fact Smalheiser & Torvik (2009) indicated that this happens when authors have more than 300 publications. In other words, although the precision of our gold standard set is 100% (basically we can assume that all are correct publications, as verified by their authors) it seems that the recall of the golden standard set is not necessary 100%. Thus, the values of 'wrong' publications obtained through our methodologies might not be as high in reality, and thus we can consider this measure to be the upper bound of false positives that we could expect for the whole analysis, because true values will likely be smaller. The methodology developed in this paper will be further applied in an impact study of the set of Dutch full professors, retrieving citations to all their publications. We would like to point out that our methodology, relying on domain specific scholarly communication, is sensitive towards the disciplinary composition of the author set, e.g., authors that publish mostly books are underrepresented. This will be explored in further analysis.

Note that the success of cross-database retrieval and author disambiguation heavily depends on access policies of the hosting institutions, and the quality of the databases involved. Even if access is given, extensive institutional collaboration is required to interlink and harmonize databases. Initiatives such as ORCID (Foley & Kochalk, 2010) with the idea of a central registry of unique identifiers for individual researchers or bottom-up networked approaches such as the VIVO international researcher network (Börner et al, 2012) that assigns unique VIVO identifiers to each scholar, aim to provide processes and data structures to assign and keep track of unique scholars and their continuously evolving oeuvres. The data collected by ORCID and VIVO can be used as additional 'gold standards' in future evaluation studies. The methodology presented here can be applied to retrieve publications for scholars with a valid ORCID and VIVO from the existing commercial and public data sources. Ultimately, unique author identifiers are required for the comprehensive analysis of science, e.g., using altmetrics (Wouters & Costas, 2012), and also for models of science (Scharnhorst et al. 2012) using data from multiple databases.

## Acknowledgements

The authors acknowledge detailed comments and suggestions provided by Vincent Lariviere and Kevin W. Boyak for an early version of this manuscript. Part of this work was funded by the NoE European InterNet Science (EINS) (EC Grant 288021) and the National Institutes of Health under award U01 GM098959.

---

[1] The idea of combining different data sources with the objective of collecting data at the individual level is not completely new (see for example D'Angelo et al, 2011) and has shown already interesting results.

[2] Exceptions exist with addresses like info@ or dep@

[3] We expect sometimes an e-mail is transferred to another researcher with the same (or very similar) name in the same organization when the previous e-mail holder has left the organization.